\documentclass[aps,prd,onecolumn,groupedaddress,showpacs,nofootinbib,amssymb]{revtex4}
\usepackage[dvips]{graphicx}
\usepackage{amssymb}
\usepackage{amsmath}
\usepackage{graphicx}
\usepackage{amsfonts}
\usepackage{bm}

\begin{document}

\title{Singular inflation from generalized equation of state fluids}
\author{
S.~Nojiri,$^{1,2}$\,\thanks{nojiri@gravity.phys.nagoya-u.ac.jp}
S.~D.~Odintsov,$^{3,4,6}$\,\thanks{odintsov@ieec.uab.es}
V.~K.~Oikonomou,$^{5,6}$\,\thanks{v.k.oikonomou1979@gmail.com}}
\affiliation{
$^{1)}$ Department of Physics, Nagoya University, Nagoya 464-8602, Japan \\
$^{2)}$ Kobayashi-Maskawa Institute for the Origin of Particles and the
Universe, Nagoya University, Nagoya 464-8602, Japan \\
$^{3)}$Institut de Ciencies de lEspai (IEEC-CSIC),
Campus UAB, Carrer de Can Magrans, s/n\\
08193 Cerdanyola del Valles, Barcelona, Spain\\
$^{4)}$ ICREA,
Passeig Llu{\^ i}s Companys, 23,
08010 Barcelona, Spain\\
$^{5)}$ Department of Theoretical Physics, Aristotle University of
Thessaloniki,
54124 Thessaloniki, Greece\\
$^{6)}$ National Research Tomsk State University, 634050 Tomsk and
Tomsk State Pedagogical University, 634061 Tomsk, Russia}

\begin{abstract}

We study models with a generalized inhomogeneous equation of state fluids, in the
context of singular inflation, focusing to so-called Type IV singular
evolution. In the simplest case, this cosmological fluid is described by
an equation of state with constant $w$, and therefore a direct modification
of this constant $w$ fluid, is achieved by using a generalized form of an
equation of state. We investigate from which models with generalized
phenomenological equation of state, a Type IV singular inflation can be
generated and what the phenomenological implications of this singularity
would be. We support our results with illustrative examples and we also
study the impact of the Type IV singularities on the slow-roll parameters
and on the observational inflationary indices, showing the consistency
with Planck mission results. The unification of singular inflation with
singular dark energy era for specific generalized fluids is also proposed.
\end{abstract}


\maketitle



\def\pp{{\, \mid \hskip -1.5mm =}}
\def\cL{\mathcal{L}}
\def\be{\begin{equation}}
\def\ee{\end{equation}}
\def\bea{\begin{eqnarray}}
\def\eea{\end{eqnarray}}
\def\tr{\mathrm{tr}\, }
\def\nn{\nonumber \\}
\def\e{\mathrm{e}}

\section{Introduction}

One of the most astonishing surprises that the scientific community
experienced in the late 90's was the observationally verified late-time
acceleration of the Universe \cite{riess}, an observation which was
contrary to any up to that date perception or expectation for the
Universe's late-time evolution. Since then, considerable amount of research
was conducted towards the understanding and explanation of this late-time
acceleration. In addition, a desirable feature that a complete theory of
cosmological evolution should have is the description of the early-time
acceleration, known as inflationary era \cite{inflation,inflationreview},
and of the late-time acceleration using the same theoretical framework. One
of the most successful approaches that describe late-time and early-time
acceleration in unified manner \cite{sergeinojiri2003} is provided by $F(R)$
theories \cite{reviews}, which serve as modifications of the standard Einstein-Hilbert gravity.
It is very interesting that modification of gravity may be considered as addition
of (geometric) terms to the phenomenological equation of state of corresponding
generalized fluid. In fact, it is quite well-known that generalized (imperfect)
fluid EoS \cite{inhomogen} which contain inhomogeneous terms may successfully describe the universe acceleration.
The homogeneous modifications assume the
inclusion of terms in the EoS which depend on the effective energy density,
and the inhomogeneous modifications assume terms that depend explicitly on
the Hubble rate or its higher derivatives, and as was demonstrated in
\cite{inhomogen}, in the context of inhomogeneous phenomenological EoS
theories, it is possible to describe even phantom evolution without introducing
a scalar field with negative kinetic energy. Apart from this
appealing feature of phenomenological EoS theories, a strong motivation to
use and study these comes from the fact that inhomogeneous terms may be
understood as the time-dependent bulk or shear viscosity \cite{brevik1}
and also symmetry considerations indicate such
modifications of EoS \cite{brevik2}.
Such generalized fluids may be used for the construction of inflationary era
(see, for instance \cite{fluid}).
Finally and most importantly, it can
be proven that in the context of modified gravity theories, the resulting
EoS of the geometric fluid is modified in the fashion we described above
\cite{reviews,nojodineos1}.

It is common in the context of these theories, however, that various types
of finite time singularities frequently occur. It is remarkable that such
finite-time singularities are universal, as they may occur after early-time acceleration
as well as after late-time acceleration. The purpose of this article
is to study the Universe's cosmological evolution using a phenomenological
EoS for the dark fluid, that leads to, or is responsible for, a Type IV
finite time singularities.
The interest to this specific singularity is owing to the fact that the Universe's
evolution continues smoothly after passing through this singularity!
This argument is strongly supported by the fact that no geodesics incompleteness
occurs for sudden and this types of singularities, so in principle the evolution is not
abruptly interrupted by the singular behavior of some physical quantities, such as
the higher derivatives of the Hubble rate. The study of realistic cosmological
singularities was initiated sometime ago \cite{barrowsing1}. The very interesting finite
time future singularities are known as sudden
singularities or Type II ones (for later studied on these, see
\cite{barrowsing2,barrow,Barrow:2015ora}). Also the finite time
cosmological singularities were firstly classified in \cite{Nojiri:2005sx},
with the classification taking into account whether physical quantities
like the scale factor, the effective energy density, or pressure, are
finite or infinite at certain spacetime points. The Type IV singularity is
the most ``mild'' singularity among the four types of singularities, with
the mild referring to the fact that these are not crushing type
singularities. For recent studies on these singularities see
\cite{noo1,noo2,noo3}, where details can be found for the various types of
cosmological evolution that these can generate or modify.

The purpose of this letter is to demonstrate at first time that singular
inflation induced by generalized fluids is quite possible. The unification
of singular inflation with singular dark energy era in the universe filled
with generalized fluid is also proposed. The paper outline is: In section II, we present some essential information for the theoretical framework we shall use and we study in detail the how a singular evolution can be produced by generalized EoS models and we also provide some concrete examples. In section III, we address the slow-roll behavior issue of these generalized EoS theories and in section IV the conclusions follow. 

\section{Theoretical framework and analysis of singular generalized EoS models}

To start with, consider a perfect fluid coupled with the standard
Einstein-Hilbert gravity, with the corresponding Friedmann-Robertson-Walker
(FRW) equations being equal to,
\be
\label{JGRG11}
\rho =\frac{3}{\kappa^2}H^2 \, ,\quad p = -\frac{1}{\kappa^2}\left(3H^2 +
2\dot H\right)\, ,
\ee
where it is assumed that the background metric is a flat FRW metric of the
following form,
\begin{equation}\label{metricformfrwhjkh}
\mathrm{d}s^2=-\mathrm{d}t^2+a^2(t)\sum_i\mathrm{d}x_i^2\, .
\end{equation}
In Eq.~(\ref{JGRG11}), $H$ denotes the usual the Hubble rate
$H(t)=\dot{a}(t)/a(t)$. For a given cosmological evolution in terms of the
Hubble rate $H=H(t)$, the right hand side of the two equations appearing in
Eq.~(\ref{JGRG11}) are solely functions of $t$,
\be
\label{C1}
\rho =f_\rho(t)\, , \quad p = f_p (t)\, .
\ee
Then, by solving the first equation with respect to the cosmic time $t$, we
get, $t=f_\rho^{-1} \left(\rho \right)$, and by substituting the resulting
expression into the second equation of Eq.~(\ref{C1}),
we obtain the functional form of the EoS, namely,
\be
\label{C2}
p = f_p \left( f_\rho^{-1} \left(\rho \right) \right)\, .
\ee
In this way we can obtain the exact form of the effective EoS, which we
denote as $w_{\mathrm{eff}}$. It is worth recalling in brief the essential
features of the homogeneous and inhomogeneous phenomenological EoS
theories, the complete presentation of which can be found in
\cite{inhomogen}. In the
context of inhomogeneous phenomenological EoS theories the EoS is modified
as follows,
\begin{equation}\label{inhoeos}
p =-\rho -f(\rho )+G(H) \, ,
\end{equation}
and when the function $G(H)$ is zero, this corresponds to the homogeneous EoS theory.
A much more general functional dependence of the effective pressure as a
function of the effective energy density and the Hubble rate, is given by
the following functional form,
\be
\label{C5}
p = f \left( \rho , H \right) \, .
\ee
In Eqs.~(\ref{inhoeos}) and (\ref{C5}), the functions $f$ and $G$ are
arbitrary functions of their arguments.

The focus in this letter is to study the phenomenological EoS theories of
Eq.~(\ref{C5}), for a Type IV cosmological evolution as it occurs
in the inflationary epoch. It is worth recalling
in brief the classification of finite time cosmological singularities, as
in Ref.~\cite{Nojiri:2005sx}:
\begin{itemize}
\item Type I (``Big Rip'') : When $t \to t_s$, the scale factor $a$, the
effective energy
density $\rho_\mathrm{eff}$, and the effective pressure $p_\mathrm{eff}$
diverge, namely
$a \to \infty$, $\rho_\mathrm{eff} \to \infty$, and
$\left|p_\mathrm{eff}\right| \to \infty$. For details on the Big Rip finite
singularity, the reader is referred to
\cite{Caldwell:2003vq,Nojiri:2005sx,ref5}.
\item Type II (``sudden''): When $t \to t_s$, although both of the scale
factor and the effective
energy density are finite, that is, $a \to a_s$, $\rho_\mathrm{eff} \to
\rho_s$, the effective pressure diverges, namely
$\left|p_\mathrm{eff}\right| \to \infty$. See \cite{barrowsing2,barrow} for
further analysis on this type of singularities.
\item Type III: When $t \to t_s$, although the scale factor is finite, $a
\to a_s$, both of the
effective energy density and the effective pressure diverge,
$\rho_\mathrm{eff} \to
\infty$, $\left|p_\mathrm{eff}\right| \to \infty$.
\item Type IV: When $t \to t_s$, all of the scale factor, the effective
energy density, and
the effective pressure are finite, that is, $a \to a_s$, $\rho_\mathrm{eff}
\to \rho_s$,
$\left|p_\mathrm{eff}\right| \to p_s$, but the higher derivatives of the
Hubble rate diverge. For a detailed account on this singularity, see
\cite{Nojiri:2005sx}.
\end{itemize}
Here the effective energy
density $\rho_\mathrm{eff}$, and the effective pressure $p_\mathrm{eff}$
are defined by $\rho_\mathrm{eff} \equiv \left(3/\kappa^2\right)
H^2$, and $p_\mathrm{eff} \equiv - \left(1/\kappa^2\right) \left( 2
\dot H + 3 H^2 \right) $.
A quite general form of the Hubble rate that can lead to Type II and Type
IV singular cosmological evolution, is the following,
\be
\label{IV1}
H(t) = f_1(t) + f_2(t) \left( t_s - t \right)^\alpha\, ,
\ee
where the functions $f_1(t)$ and $f_2(t)$ are considered to be smooth and
differentiable functions of $t$. The Type II singularity occurs when the
constant parameter $\alpha$ is restricted to take the values, $0<\alpha<1$,
while when $\alpha>1$, the cosmological evolution develops a Type IV
singularity. In the following, and in order to avoid inconsistencies, we
assume that the parameter $\alpha $ appearing in Eq.~(\ref{IV1}), takes the
following form,
\be
\label{IV2}
\alpha= \frac{n}{2m + 1}\, ,
\ee
where $n$ and $m$ are arbitrarily chosen positive integers. In addition, a
modified version of the Hubble rate appearing in Eq.~(\ref{IV1}), is given
below,
\be
\label{IV1B}
H(t) = f_1(t) + f_2(t) \left| t_s - t \right|^\alpha\, ,
\ee
in which case, $\alpha$ can take more general values, without assuming the
restricted form given in Eq.~(\ref{IV2}), but we study here only the case
for which $\alpha $ has the form given in Eq.~(\ref{IV2}). In the following
we shall investigate which homogeneous and inhomogeneous EoS can generate
the cosmological evolution appearing in Eq.~(\ref{IV1}), with special
emphasis given in the Type IV singularity. In general, it is difficult to
find an explicit form of the EoS, for complicated forms of the functions
$f_1(t)$ and $f_2(t)$, so let us start with a simple example describing a
Type IV evolution, with the functions $f_1(t)$ and $f_2(t)$ being chosen
as, $f_1(t)=0$ and $f_2(t) = f_0$, where $f_0$ is an arbitrary positive
constant. By using these values and by substituting in Eq.~(\ref{JGRG11})
the Hubble rate of Eq.~(\ref{IV1}), the effective energy density $\rho $
and the effective pressure $p $ read,
\be
\label{C3}
\rho =\frac{3f_0^2 }{\kappa^2}\left( t_s - t \right)^{2\alpha} \, ,\quad p
= -\frac{1}{\kappa^2}\left( 3f_0^2 \left( t_s - t \right)^{2\alpha}
+ 2\alpha f_0 \left( t_s - t \right)^{\alpha -1}\right)\, ,
\ee
By using the first equation of Eq.~(\ref{C3}), we can solve it in terms of
$t_s-t$ and therefore by substituting in the effective pressure expression,
we obtain in explicit form of the EoS, which reads (see also
\cite{Nojiri:2005sx}),
\be
\label{C4}
p = - \rho - 2 \cdot 3^{- \frac{\alpha - 1}{2\alpha}} \alpha \kappa^{-
\frac{\alpha + 1}{\alpha}} f_0^{2 - \alpha} \rho ^\frac{\alpha -
1}{2\alpha}\, .
\ee
The exact type of the finite time singularities that may appear in this
cosmological evolution may be found by using the explicit forms of the
effective energy density $\rho $ and of the effective pressure $p $, as
these appear in Eq.~(\ref{C3}). By introducing the parameter
$\tilde\alpha\equiv \frac{\alpha - 1}{2\alpha}$, the cosmological evolution
of the phenomenological EoS (\ref{C4}), as described by the energy density
and pressure appearing in Eq.(\ref{C3}), has the following singularity
structure,
\begin{itemize}
\item In the case that $\tilde\alpha>1$, or in terms of $\alpha$,
$-1<\alpha<-\frac{1}{2}$, a Type III singularity occurs.
\item In the case that $\frac{1}{2}<\tilde\alpha<1$, that is, $\alpha<-1$,
a Type I singularity appears.
\item In the case that $0<\tilde\alpha<\frac{1}{2}$, or equivalently,
$\alpha>1$, a Type IV singularity occurs.
\item In the case that $\tilde\alpha<0$, or in terms of $\alpha$,
$-\frac{1}{2}<\alpha<1$, a Type II singularity occurs.
\end{itemize}
The form of the EoS appearing in Eq.~(\ref{C4}), can be viewed as a
homogeneous phenomenological EoS theory of the form (\ref{inhoeos}), with
$f(\rho )$ being equal to,
\begin{equation}\label{view1}
f(\rho )=- 2 \cdot 3^{- \frac{\alpha - 1}{2\alpha}} \alpha \kappa^{-
\frac{\alpha + 1}{\alpha}} f_0^{2 - \alpha} \rho ^\frac{\alpha -
1}{2\alpha}\, .
\end{equation}
and $G(H)=0$. In addition, the EoS appearing in Eq.~(\ref{C4}) can also be viewed as an
inhomogeneous phenomenological EoS theory of the form (\ref{inhoeos}), by
using the fact that $\rho $ can be written in terms of the Hubble rate
$H^2$ in the way dictated by Eq.~(\ref{JGRG11}). By doing so, the EoS can
be written in the following form,
\be
\label{C6}
p = - \rho - \frac{2 \alpha}{\kappa^2} f_0^{2 - \alpha} H^\frac{\alpha -
1}{\alpha}\, .
\ee
which is of the form given in Eq.~(\ref{inhoeos}), with $f(\rho )=0$ and
with $G(H)$ being equal to,
\begin{equation}\label{gh}
G(H)=- \frac{2 \alpha}{\kappa^2} f_0^{2 - \alpha} H^\frac{\alpha -
1}{\alpha} \, .
\end{equation}
We may also consider the following model,
\be
\label{C9}
H(t) = h_0 \left\{ \left( \frac{t - t_0}{t_1} \right)^{-2n} + 1
\right\}^{-\frac{\alpha}{2n}}\, ,
\ee
where $h_0$, $t_0$, $t_1$, $\alpha$ are constants and we assume $h_0>0$,
$n>0$ and also that 
$0<\alpha<1$. 
$\alpha>1$. 
When $t\to \pm \infty$, we find $H(t)$
becomes a constant $H(t) \to h_0$ and when $t\sim t_0$, a Type IV
singularity occurs, since the Hubble rate behaves as,
$H(t) \sim h_0 \left( \frac{t - t_0}{t_1} \right)^\alpha$. The
corresponding effective energy density and effective pressure for the
Hubble rate (\ref{C9}) are equal to,
\begin{align}
\label{C9b}
\rho =& \frac{3 h_0^2}{\kappa^2} \left\{ \left( \frac{t - t_0}{t_1}
\right)^{-2n} + 1 \right\}^{-\frac{\alpha}{n}}\, , \nn
p =& - \frac{3 h_0^2}{\kappa^2} \left\{ \left( \frac{t - t_0}{t_1}
\right)^{-2n} + 1 \right\}^{-\frac{\alpha}{n}}
- \frac{2\alpha h_0}{\kappa^2 t_1} \left\{ \left( \frac{t - t_0}{t_1}
\right)^{-2n} + 1 \right\}^{-\frac{\alpha}{2n} - 1}
\left( \frac{t - t_0}{t_1} \right)^{-2n-1}\, .
\end{align}
Consequently, by making use of Eq.~(\ref{C2}), we obtain the following EoS,
\be
\label{C9c}
p = - \rho - \frac{2\alpha h_0}{\kappa^2 t_1} \left( \frac{\kappa^2 \rho
}{3 h_0^2} \right)^{\frac{1}{2} + \frac{n}{\alpha}}
\left\{ \left( \frac{\kappa^2 \rho }{3 h_0^2} \right)^{- \frac{n}{\alpha}}
- 1 \right\}^{1 + \frac{1}{2n}}\, .
\ee
Now consider the case in which the Hubble rate is given by,
\begin{equation}\label{hubnew}
H(t)=f_0 (t-t_1)^{\alpha } +c_0(t-t_2)^{\beta }\, ,
\end{equation}
with $c_0$ and $f_0$ constant and positive parameters, and $\alpha$,~$\beta
>1$, so that the cosmological evolution has two Type IV singularities at
$t=t_1$ and $t=t_2$. We may choose $t_1$ to be at the end of the
inflationary era and $t_2$ to be at late-time. In principle, it is quite
difficult to obtain the exact form of the EoS for the Hubble rate
(\ref{hubnew}), since it is quite difficult to solve the equation $\rho
\sim H^2(t)$ with respect to $t$. However we can find an approximate form
of the EoS, near the two Type IV singularities. Before going into the
details of this approximation, we quote here the cosmic time dependence of
the effective energy and of the effective pressure for the Hubble rate
(\ref{hubnew}), which are,
\begin{align}
\label{energypress}
\rho =&\frac{3 f_0^2 (t-t_1)^{2 \alpha }}{\kappa ^2}+\frac{6 c_0 f_0
(t-t_1)^{\alpha } (t-t_2)^{\beta }}{\kappa ^2}+\frac{3 c_0^2 (t-t_2)^{2
\beta }}{\kappa ^2} \, ,\nn
p =&-\frac{3 f_0^2 (t-t_1)^{2 \alpha }}{\kappa ^2}-\frac{6 c_0 f_0
(t-t_1)^{\alpha } (t-t_2)^{\beta }}{\kappa ^2}-\frac{3 c_0^2 (t-t_2)^{2
\beta }}{\kappa ^2}-\frac{2 f_0 (t-t_1)^{-1+\alpha } \alpha }{\kappa
^2}-\frac{2 c_0 (t-t_2)^{-1+\beta } \beta }{\kappa ^2}\, .
\end{align}
So at the vicinity of the early-time Type IV singularity, these read,
\begin{align}\label{energypress2}
& \rho =\frac{3 c_0^2 (t-t_2)^{2 \beta }}{\kappa ^2}\, , \quad p =-\frac{3
c_0^2 (t-t_2)^{2 \beta }}{\kappa ^2}-\frac{2 c_0 (t-t_2)^{-1+\beta } \beta
}{\kappa ^2} \, ,
\end{align}
so the EoS takes the approximate form,
\begin{equation}\label{neweqna}
p =-\rho -\frac{2 c_0\beta }{\kappa ^2} \left(\frac{\rho \kappa ^2}{3
c_0^2}\right)^{\frac{\beta -1}{2 \beta }}\, .
\end{equation}
The resulting physical situation is quite appealing, since it seems that
the late-time singularity controls the early-time EoS for the inhomogeneous
phenomenological EoS theory that generates the Hubble rate (\ref{hubnew}),
near the of course early-time singularity. In fact, the resulting
inhomogeneous phenomenological EoS near the early-time singularity is
controlled solely from the late-time singularity, since the terms $\sim
(t-t_1)$ vanish for a Type IV early-time singularity. Of course, for other
types of singularities, this may not occur, so only the Type IV case has
this interesting feature. The same applies for the late-time singularity,
so near the late-time Type IV singularity the effective energy density and
effective pressure are,
\begin{align}\label{energypress1}
& \rho =\frac{3 c_0^2 (t-t_1)^{2 \alpha }}{\kappa ^2}\, , \quad p
=-\frac{3 c_0^2 (t-t_1)^{2 \alpha }}{\kappa ^2}-\frac{2 c_0
(t-t_1)^{-1+\alpha } \alpha }{\kappa ^2}\, ,
\end{align}
so that the corresponding EoS reads,
\begin{equation}\label{neweqna1}
p =-\rho -\frac{2 c_0\alpha }{\kappa ^2} \left(\frac{\rho \kappa ^2}{3
c_0^2}\right)^{\frac{\alpha -1}{2 \alpha }}\, .
\end{equation}
Therefore, the resulting picture is that the EoS near the late-time
singularity is solely controlled from the early-time Type IV singularity,
which is a quite interesting feature. So by suitably choosing the
parameters, this effect can be quite large or even negligible. For a
relevant study in which this phenomenology also occurs, see also
\cite{noo3}.


Another interesting model with the property that it provides a unified
description of inflation at early-time with a late-time acceleration
evolution with a Type IV singularity occurring at late-time. The Hubble
rate of this model is given below,
\be
\label{hubnoj1}
H(t)=\frac{f_1}{\sqrt{t^2+t_0^2}}+\frac{f_2 t^2 (-t+t_1)^{\alpha
}}{t^4+t_0^4}+f_3 (-t+t_2)^{\beta } \, .
\ee
Notice that the late-time singularity can even occur at present time. For a
detailed account on the consequences of this Type IV singularity occurring
at present time, see \cite{noo1}. In Eq.~(\ref{hubnoj1}), the parameters
$\alpha$, $\beta$, $t_0$, $f_1$, $f_2$, and $f_3$ are chosen to be positive
constants so that we ensure that $H(t)>0$. 
Then, 
the effective energy density $\rho $ and the effective pressure $p $ are
equal to,
\begin{align}
\label{C8}
\rho = & \frac{3}{\kappa^2}
\left\{ \frac{f_1}{\sqrt{t^2+t_0^2}}+\frac{f_2 t^2 (-t+t_1)^{\alpha
}}{t^4+t_0^4}+f_3 (-t+t_2)^{\beta } \right\}^2 \, , \nn
p =& - \frac{3}{\kappa^2}
\left\{ \frac{f_1}{\sqrt{t^2+t_0^2}}+\frac{f_2 t^2 (-t+t_1)^{\alpha
}}{t^4+t_0^4}+f_3 (-t+t_2)^{\beta } \right\}^2 \nn
& - \frac{2}{\kappa^2} \left\{ -\frac{f_1
t}{\left(t^2+t_0^2\right)^{3/2}}-\frac{4 f_2 t^5 (-t+t_1)^{\alpha
}}{\left(t^4+t_0^4\right)^2}+\frac{2 f_2 t (-t+t_1)^{\alpha }}{t^4+t_0^4}
\right. \nn
& \left. -\frac{f_2 t^2 (-t+t_1)^{-1+\alpha } \alpha }{t^4+t_0^4} - f_3
\beta(-t+t_2)^{-1+\beta } \right\} \, .
\end{align}
The cosmic dark fluid scenario with effective pressure and energy density
as in Eq.~(\ref{C8}) can describe a plethora of cosmological evolutions.
For example, $t_1$ may correspond to early-time and $t_2$ at late-time, so
if $\alpha>1$ and $\beta>1$, a Type IV singularity occurs at both early-
and late-time, as in the previous example. Let us briefly repeat the same
analysis as in the previous case, in order to find an analytic
approximation for the EoS near the singularities. Consider first the case
that $t\simeq t_1$, so the physical system is considered to be near the
early-time Type IV singularity. Then, the effective energy density and
pressure become approximately equal to,
\begin{align}
\label{C8new}
\rho \simeq & \frac{3 f_1^2}{\left(t^2+t_0^2\right) \kappa ^2}+\frac{6 f_1
f_3 (-t+t_2)^{\beta }}{\sqrt{t^2+t_0^2} \kappa ^2}+\frac{3 f_3^2
(-t+t_2)^{2 \beta }}{\kappa ^2} \, , \nn
p \simeq& \frac{2 f_1 t}{\left(t^2+t_0^2\right)^{3/2} \kappa ^2}-\frac{3
f_1^2}{\left(t^2+t_0^2\right) \kappa ^2}-\frac{6 f_1 f_3 (-t+t_2)^{\beta
}}{\sqrt{t^2+t_0^2} \kappa ^2}-\frac{3 f_3^2 (-t+t_2)^{2 \beta }}{\kappa
^2}+\frac{2 f_3 (-t+t_2)^{-1+\beta } \beta }{\kappa ^2} \, .
\end{align}
and as is obvious, the late-time Type IV singularity controls the effective
energy density and effective pressure of the cosmological dark fluid, near
the early-time Type IV singularity. In order to see how the late-time
singularity controls the EoS of the phenomenological theory, let us further
simplify the expressions appearing in Eq.~(\ref{C8new}), by taking into
account that $t\ll t_2$, in which case the effective energy density and
effective pressure read,
\begin{align}
\label{C8newnew}
\rho \simeq \frac{3 f_3^2 (-t+t_2)^{2 \beta }}{\kappa ^2}\simeq \frac{3
f_3^2 t_2^{2 \beta }}{\kappa ^2}\, , \quad
p \simeq -\frac{3 f_3^2 (-t+t_2)^{2 \beta }}{\kappa ^2}\simeq -\frac{3
f_3^2 t_2^{2 \beta }}{\kappa ^2} \, .
\end{align}
so that the EoS reads, $p \simeq -\rho $. Also notice that, owing to the
appearance of the parameter $t_2$, the late-time singularity drastically
affects the early-time EoS and acceleration. The latter, is due to the fact
that the early-time evolution is nearly de Sitter acceleration (since $p
\simeq -\rho $).

Conversely, at late-time and near the future Type IV singularity, that is, 
when $t\simeq t_2$, the corresponding effective energy density and pressure
of the phenomenological EoS theory are,
\begin{align}
\label{C8new1}
\rho \simeq & \frac{3 f_1^2}{\left(t^2+t_0^2\right) \kappa ^2}+\frac{6 f_0
f_1 t^2 (-t+t_1)^{\alpha }}{\sqrt{t^2+t_0^2} \left(t^4+t_0^4\right) \kappa
^2}+\frac{3 f_0^2 t^4 (-t+t_1)^{2 \alpha }}{\left(t^4+t_0^4\right)^2 \kappa
^2} \, , \nn
p \simeq& \frac{2 f_1 t}{\left(t^2+t_0^2\right)^{3/2} \kappa ^2}-\frac{3
f_1^2}{\left(t^2+t_0^2\right) \kappa ^2}+\frac{8 f_0 t^5 (-t+t_1)^{\alpha
}}{\left(t^4+t_0^4\right)^2 \kappa ^2}-\frac{4 f_0 t (-t+t_1)^{\alpha
}}{\left(t^4+t_0^4\right) \kappa ^2}-\frac{6 f_0 f_1 t^2 (-t+t_1)^{\alpha
}}{\sqrt{t^2+t_0^2} \left(t^4+t_0^4\right) \kappa ^2} \nn & -\frac{3 f_0^2
t^4 (-t+t_1)^{2 \alpha }}{\left(t^4+t_0^4\right)^2 \kappa ^2}+\frac{2 f_0
t^2 (-t+t_1)^{-1+\alpha } \alpha }{\left(t^4+t_0^4\right) \kappa ^2} \, ,
\end{align}
so practically, we could say that the early-time Type IV singularity
controls the late-time behavior of the phenomenological EoS theory. Let us
see however the extent of this control explicitly. Since we assumed that
$t\gg t_1$, the effective energy density and pressure appearing in
Eq.~(\ref{C8new1}) are very much simplified, and by keeping only leading
order terms, these become approximately equal to,
\begin{equation}\label{apprxlatetime}
\rho \simeq \frac{3 f_0^2 t^4 (-t+t_1)^{2 \alpha
}}{\left(t^4+t_0^4\right)^2 \kappa ^2}\, , \quad p \simeq -\frac{3 f_0^2
t^4 (-t+t_1)^{2 \alpha }}{\left(t^4+t_0^4\right)^2 \kappa ^2}\, ,
\end{equation}
and hence it easily follows that in this case, the EoS is $p \simeq -\rho
$, so we have late-time de Sitter acceleration with $w_{\mathrm{eff}}\simeq
-1$. Therefore, the contribution of the early-time singularity is not so
important in this case. Another interesting example with two different Type
IV singularities is described by the following cosmological evolution,
\begin{equation}\label{hnew1}
H(t)=f_0+c \left(t-t_1\right)^{\alpha}\left(t-t_2\right)^{\beta}\, ,
\end{equation}
in which case, if $\alpha,\beta>1$, Type IV singularities occur at $t=t_1$
and at $t=t_2$. The corresponding effective energy density and pressure are
equal to,
\begin{align}\label{effpressandensity}
\rho =& \frac{3 c^2}{\kappa ^2}+\frac{6 c f_0 (-t+t_1)^{\alpha }
(-t+t_2)^{\beta }}{\kappa ^2}+\frac{3 f_0^2 (-t+t_1)^{2 \alpha }
(-t+t_2)^{2 \beta }}{\kappa ^2} \, , \nn
p =&-\frac{3 c^2}{\kappa ^2}-\frac{6 c f_0 (-t+t_1)^{\alpha }
(-t+t_2)^{\beta }}{\kappa ^2}-\frac{3 f_0^2 (-t+t_1)^{2 \alpha }
(-t+t_2)^{2 \beta }}{\kappa ^2} \nn &
+\frac{2 f_0 (-t+t_1)^{-1+\alpha } (-t+t_2)^{\beta } \alpha }{\kappa
^2}+\frac{2 f_0 (-t+t_1)^{\alpha } (-t+t_2)^{-1+\beta } \beta }{\kappa
^2}\, .
\end{align}
At both singular points, the effective energy and pressure are exactly
equal to,
\begin{equation}\label{exactexpress}
\rho =\frac{3 c^2}{\kappa ^2} \, , \quad p =-\frac{3 c^2}{\kappa ^2}\, ,
\end{equation}
which implies that the EoS is exactly equal to minus one, since $p =-\rho
$. We have to note that in all the above paradigms, the results can
drastically change, if the singularities we considered are not Type IV.


The converse procedure is possible to produce a Type IV singular
cosmological evolution, from a given phenomenological EoS theory. Indeed,
consider a homogeneous phenomenological EoS theory, with the EoS being of
the form,
\begin{equation}\label{pesoparadigmfinal}
p =-\rho +f(\rho )\, ,
\end{equation}
with $f(\rho )$ being equal to,
\begin{equation}\label{reqt}
f(\rho )=A \rho ^{\alpha} \, .
\end{equation}
Then, by using the energy conservation law, we easily obtain that,
\begin{equation}\label{testeqn}
\rho =(t-t_0)^{\frac{2}{1-\alpha}}\left(\frac{\sqrt{3}\kappa
A}{2}\right)^{\frac{2}{1-\alpha}}\, .
\end{equation}
Correspondingly, the effective pressure as a function of time reads,
\begin{align}\label{presstest}
p = &-\frac{\left(\frac{9}{16}\right)^{\frac{\alpha }{1-2 \alpha }}
(t-t_0)^{\frac{4 \alpha }{1-2 \alpha }} (A \kappa )^{\frac{2}{1-2 \alpha }}
\left(\left(\frac{3}{4}\right)^{\frac{1}{1-2 \alpha }}
(t-t_0)^{\frac{2}{1-2 \alpha }} (A \kappa )^{\frac{2}{1-2 \alpha
}}\right)^{-2 \alpha }}{(1-2 \alpha )^2 \kappa ^2} \nn &
\times \frac{\left(\left(\frac{3}{4}\right)^{\frac{1}{1-2 \alpha }}
(t-t_0)^{\frac{2}{1-2 \alpha }} (A \kappa )^{\frac{2}{1-2 \alpha
}}+\left(\left(\frac{3}{4}\right)^{\frac{1}{1-2 \alpha }}
(t-t_0)^{\frac{2}{1-2 \alpha }} (A \kappa )^{\frac{2}{1-2 \alpha
}}\right)^{\alpha }\right)}{(1-2 \alpha )^2 \kappa ^2}\, ,
\end{align}
so when $0<\alpha <\frac{1}{2}$, a Type IV singularity occurs in the
cosmological evolution, at $t=t_0$.

\section{Slow-roll parameters for singular generalized EoS models}

In all the above cases, we investigated the general behavior of various
homogeneous or inhomogeneous phenomenological EoS theories, in the presence
of a Type IV singularity during the cosmological evolution. However, as was
stressed in \cite{noo1}, the presence of a Type IV singularity during the
inflationary era may have dramatic consequences on the inflationary
observational indices. It is therefore compelling to study the behavior of
these observational indices in the presence of a Type IV singularity. We
shall stay at a qualitative level analysis, but we shall attempt to compare
some results with the recent Planck observational data \cite{planck}. This
analysis however is strongly model dependent, so in this letter we will
just highlight the most important qualitative implications of the Type IV
singularities.

The full analysis on the derivation of the slow-roll parameters for a
perfect fluid phenomenological EoS theory can be found in 
Refs.~\cite{nojodineos1}, and the key point is that the FRW equations
appearing in Eq.~(\ref{JGRG11}) can be expressed in terms of the
$e$-folding number $N$, as follows,
\begin{equation}\label{efold1}
\rho =\frac{3}{\kappa^2}\left(H(N)\right)^2 \, ,\quad p (N)+\rho (N)=
-\frac{2 H(N)H'(N)}{\kappa^2}\, ,
\end{equation}
where $H'(N)=\mathrm{d}H/\mathrm{d}N$. Assuming that the EoS has the
following general form,
\be
\label{SS2}
p (N) = - \rho _\mathrm{matter}(N) + \tilde f (\rho (N))\, ,
\ee
the second equation in (\ref{efold1}), takes the following form,
\begin{equation}\label{efold2}
\tilde f (\rho (N))= -\frac{2 H(N)H'(N)}{\kappa^2}\, .
\end{equation}
The energy density and the pressure satisfy the conservation law,
\begin{equation}\label{conlaw}
\rho' (N)+3H(N)\left( \rho (N)+p
(N)\right)=0\, ,
\end{equation}
where $\rho' (N)=\mathrm{d}\tilde f (\rho (N))/\mathrm{d}N$
and in view of Eq.~(\ref{efold2}), this conservation law becomes,
\begin{equation}\label{efold3}
\rho'(N)+3\tilde f (\rho (N))=0\, .
\end{equation}
Combining Eqs.~(\ref{efold3}) and (\ref{SS2}), we obtain,
\begin{equation}\label{preinf}
\frac{2}{\kappa^2}\left[(H'(N))^2+H(N)+H''(N)\right]=3\tilde f'(\rho
)f(\rho ) \, ,
\end{equation}
where in this case $\tilde f '(\rho (N)$ stands for $\tilde f '(\rho
(N)=\mathrm{d}\tilde f (\rho )/\mathrm{d}\rho $. Consequently, we may
express the slow-roll parameters as functions only of $\rho (N)$ and
$\tilde f(\rho (N))$ or equivalently as functions of $H(N)$ and it's
derivatives (for details see \cite{nojodineos1}). For the
present purposes we shall use the slow-roll parameters as functions of
$H(N)$, but later on we shall use the explicit form of the function $\tilde
f(\rho (N))$, in order to have comparison with observational data coming
from Planck \cite{planck}. The slow-roll parameters $\epsilon$, $\eta$ and
$\xi$ are given in terms of $H(N)$ as follows,
\begin{align}
\label{S7}
\epsilon =& - \frac{H(N)}{4 H'(N)} \left[ \frac{\frac{6 H'(N)}{H(N)}
+ \frac{H''(N)}{H(\phi)} + \left( \frac{H'(N)}{H(N)} \right)^2}
{3 + \frac{H'(N)}{H(N)}} \right]^2 \, , \nn
\eta = & -\frac{1}{2} \left( 3 + \frac{H'(N)}{H(N)} \right)^{-1} \left[
\frac{9 H'(N)}{H(N)} + \frac{3 H''(N)}{H(N)} + \frac{1}{2} \left(
\frac{H'(N)}{H(N)} \right)^2 -\frac{1}{2} \left( \frac{H''(N)}{H'(N)}
\right)^2
+ \frac{3 H''(N)}{H'(N)} + \frac{H'''(N)}{H'(N)} \right] \, , \nn
\xi^2 = & \frac{ \frac{6 H'(N)}{H(N)} + \frac{H''(N)}{H(N)}
+ \left( \frac{H'(N)}{H(N)} \right)^2 }{4 \left( 3 + \frac{H'(N)}{H(N)}
\right)^2}
\left[ \frac{3 H(N) H'''(N)}{H'(N)^2} + \frac{9 H'(N)}{H(N)}
- \frac{2 H(N) H''(N) H'''(N)}{H'(N)^3} + \frac{4 H''(N)}{H(N)}
\right.
\nn
& \left.
+ \frac{H(N) H''(N)^3}{H'(N)^4} + \frac{5 H'''(N)}{H'(N)} - \frac{3 H(N)
H''(N)^2}{H'(N)^3} - \left( \frac{H''(N)}{H'(N)} \right)^2
+ \frac{15 H''(N)}{H'(N)}
+ \frac{H(N) H''''(N)}{H'(N)^2} \right]\, .
\end{align}
It is convenient to have the slow-roll parameters as explicit functions of
the cosmic time, in order to examine the various forms of the Hubble rate
we presented in this letter, that actually lead to a Type IV singularity.
In this way, we will explicitly study how the singularity that occurs at a
finite time $t_s$, affects the slow-roll parameters and therefore the
approximation itself. 
These expressions are given below,
\begin{align}
\label{S9}
\epsilon = & - \frac{H^2}{4 \dot H} \left( \frac{6\dot H}{H^2} +
\frac{\ddot H}{H^3} \right)^2 \left( 3 + \frac{\dot H}{H^2} \right)^{-2}\,
, \nn
\eta = & - \frac{1}{2} \left( 3 + \frac{\dot H}{H^2} \right)^{-1} \left(
\frac{6\dot H}{H^2} + \frac{{\dot H}^2}{2 H^4} - \frac{\ddot H}{H^3}
- \frac{{\dot H}^4}{2 H^4} + \frac{{\dot H}^2 \ddot H}{H^5} - \frac{{\ddot
H}^2}{2H^2} + \frac{3 \ddot H}{H \dot H}
+ \frac{\dddot H}{H^2 \dot H} \right)\, ,\nn
\xi^2 = & \frac{1}{4} \left( \frac{6\dot{H}}{H^2} + \frac{\ddot{H}}{H^3}
\right) \left( 3 + \frac{\dot{H}}{H^2} \right)^{-1} \left(
\frac{9\ddot{H}}{H {\dot H} }
+ \frac{3\dddot{H}}{{\dot{H}}^2} + \frac{2 \dddot{H}}{H^2 \dot{H}} +
\frac{4 {\ddot{H}}^2}{H^2 {\dot{H}}^2}
- \frac{\ddot{H} \dddot{H}}{H {\dot{H}}^3} -
\frac{3{\ddot{H}}^2}{{\dot{H}}^3}
+ \frac{{\ddot{H}}^3}{H {\dot{H}}^4} + \frac{\ddddot{H}}{H {\dot{H}}^2}
\right)\, .
\end{align}
Assume for the moment that the Hubble rate $H(t)$ is given by the very
general form of Eq.~(\ref{IV1}), and therefore when $\alpha>1$, a Type IV
singularity is realized. Also assume that the function $f_1(t)$ is
constrained in such a way so that $f_1(t_s)$, $f_1'(t_s)$, and $f_1''(t_s)$
do not vanish. Then, the slow-roll parameters at the vicinity of the Type
IV singularity $t\sim t_s$, behave as follows,
\begin{align}
\label{S10}
\epsilon \sim & \left\{
\begin{array}{ll}
- \frac{f_1(t_s)^2}{4 \dot f_1(t_s)} \left( \frac{6\dot f_1(t_s) f_1(t_s)}{
f_1(t_s)^2} + \frac{\ddot f_1(t_s)}{ f_1(t_s)^3} \right)^2 \left( 3 +
\frac{\dot f_1(t_s)}{ f_1(t_s)^2} \right)^{-2}\, ,
& \mbox{when $\alpha>2$} \\
- \frac{f_1(t_s)^2}{4 \dot f_1(t_s)} f_2(t_s) \alpha (\alpha - 1) \left(
t_s - t \right)^{\alpha-2} \left( 3 + \frac{\dot f_1(t_s)}{ f_1(t_s)^2}
\right)^{-2}\, ,
& \mbox{when $2>\alpha>1$}
\end{array}
\right. \, , \nn
\eta \sim & \left\{
\begin{array}{ll}
- \frac{1}{2} \left( 3 + \frac{\dot f_1(t_s)}{ f_1(t_s)^2} \right)^{-1} &
\\
\quad \times \left( \frac{6\dot f_1(t_s)}{f_1(t_s)} + \frac{{\dot
f_1(t_s)}^2}{2 f_1(t_s)^4} - \frac{\ddot f_1(t_s)}{f_1(t_s)^3}
- \frac{{\dot f_1(t_s)}^4}{2 f_1(t_s)^4} + \frac{{\dot f_1(t_s)}^2 \ddot
f_1(t_s)}{f_1(t_s)^5} - \frac{{\ddot f_1(t_s)}^2}{2f_1(t_s)^2} + \frac{3
\ddot f_1(t_s)}{f_1(t_s) \dot f_1(t_s)}
+ \frac{\dddot f_1(t_s)}{f_1(t_s)^2 \dot f_1(t_s)} \right)\, ,
& \mbox{when $\alpha>3$} \\
- \frac{1}{2} \left( 3 + \frac{\dot f_1(t_s)}{ f_1(t_s)^2} \right)^{-1}
\frac{f_2 \alpha ( \alpha - 1 ) (\alpha -2 )}{f_1(t_s)^2 \dot f_1
(t_s)}\left(t_s - t \right)^{\alpha - 3}\, ,
& \mbox{when $3>\alpha>1$}
\end{array}
\right. \, , \nn
\xi^2 \sim & \left\{
\begin{array}{ll}
\frac{1}{4} \left(\frac{6\dot f_1(t_s)}{f_1(t_s)^2} + \frac{\ddot
f_1(t_s)}{f_1(t_s)^3} \right) \left( 3 + \frac{\dot f_1(t_s)}{f_1(t_s)^2}
\right)^{-1} & \\
\quad \times \left( \frac{9\ddot f_1(t_s)}{f_1(t_s) \dot f_1(t_s)}
+ \frac{3\dddot f_1(t_s)}{{\dot f_1(t_s)}^2} + \frac{2 \dddot
f_1(t_s)}{f_1(t_s)^2 \dot f_1(t_s)} + \frac{4 {\ddot
f_1(t_s)}^2}{f_1(t_s)^2 {\dot f_1(t_s)}^2}
- \frac{\ddot f_1(t_s) \dddot f_1(t_s)}{f_1(t_s) {\dot f_1(t_s)}^3}
\right. & \\
\qquad \left. - \frac{3{\ddot f_1(t_s)}^2}{{\dot f_1(t_s)}^3}
+ \frac{{\ddot f_1(t_s)}^3}{f_1(t_s) {\dot f_1(t_s)}^4} + \frac{\ddddot
f_1(t_s)}{f_1(t_s) {\dot f_1(t_s)}^2} \right)\, ,
& \mbox{when $\alpha>4$} \\
\frac{1}{4} \left( \frac{6\dot f_1(t_s)}{f_1(t_s)^2} + \frac{\ddot
f_1(t_s)}{f_1(t_s)^3} \right) \left( 3 + \frac{\dot f_1(t_s)}{f_1(t_s)^2}
\right)^{-1}
\frac{f_2(t_s) \alpha ( \alpha - 1) ( \alpha - 2) (\alpha - 3)}{f_1(t_s)
{\dot f_1(t_s)}^2} \left( t_s - t \right)^{\alpha -4}\, ,
& \mbox{when $4>\alpha>2$} \\
\frac{1}{4} \left( 3 + \frac{\dot f_1(t_s)}{f_1(t_s)^2} \right)^{-1}
\frac{f_2(t_s)^2 \alpha^2 ( \alpha - 1)^2 ( \alpha - 2) (\alpha -
3)}{f_1(t_s)^4 {\dot f_1(t_s)}^2} \left( t_s - t \right)^{2\alpha -6}\, ,
& \mbox{when $2>\alpha>1$} \\
\end{array}
\right. \, .
\end{align}
If $f_1(t)$ is assumed to be smooth, the slow-roll parameter $\epsilon$
blows up when $2>\alpha >1$, which means that it develops a singularity,
while it is regular for $\alpha>2$. In addition, the slow-roll parameter
$\eta$, blows up for $3>\alpha >1$, while $\xi^2$ blows up in two different
ways when $2>\alpha >1$ and when $4>\alpha >2$. Therefore, when $\alpha>4$, the slow-roll parameters contain no singularities at least in the vicinity of the Type IV
singularity. It is worth presenting some illustrative examples in order to
further scrutinize the behavior of the slow-roll parameters. In the case
that $H(t)=f_0\left(t-t_s\right)^{\alpha}$, the slow-roll parameter
$\epsilon$, becomes equal to,
\begin{equation}\label{slwer}
\epsilon=\frac{f_0 (t-t_s)^{-1+\alpha } \alpha (-1+6 t-6 t_s+\alpha )^2}{4
\left(3 f_0 (t-t_s)^{1+\alpha }+\alpha \right)^2}\, ,
\end{equation}
while the slow-roll parameter $\eta$ is equal to,
\begin{align}\label{sleta}
\eta =& \frac{(t-t_s)^{-3-\alpha } \left(-4 t^2+8 t t_s-4 t_s^2+4 t^2
\alpha -8 t t_s \alpha +4 t_s^2 \alpha -t^2 \alpha ^2+2 t t_s \alpha
^2\right)}{4 f_0 \left(3 f_0 (t-t_s)^{1+\alpha }+\alpha \right)}\nn &
+\frac{(t-t_s)^{-3-\alpha }\left(-t_s^2 \alpha ^2+2 \alpha ^3-2 \alpha ^4-6
f_0 (t-t_s)^{3+\alpha } (-1+3 \alpha )+f_0^2 (t-t_s)^{2 \alpha } \alpha ^2
(1+2 (-1+\alpha ) \alpha )\right)}{4 f_0 \left(3 f_0 (t-t_s)^{1+\alpha
}+\alpha \right)}{\,}.
\end{align}
Finally, the slow-roll parameter $\xi^2$ as a function of the cosmic time
$t$, is equal to,
\begin{align}\label{slwxi}
\xi^2=&\frac{(t-t_s)^{-5-2 \alpha } (-1+\alpha ) (-1+6 t-6 t_s+\alpha ) }{4
f_0^2 \left(3 f_0 (t-t_s)^{1+\alpha }+\alpha \right)}\nn &
\times \left(5 (t-t_s)^2 (-1+\alpha )^2+3 f_0^2 (t-t_s)^{2 \alpha }
(-2+\alpha )^2 (-1+\alpha ) \alpha +3 f_0 (t-t_s)^{3+\alpha } (1+2 \alpha
)\right){\,}.
\end{align}
By looking Eqs.~(\ref{slwer}), (\ref{sleta}), and (\ref{slwxi}), we can
easily see that the slow-roll parameters have singularities at the point
$t=t_s$, where the Type IV singularity occurs, a fact that we also stressed
in the general example we presented earlier. The singularity in the
slow-roll parameters can be viewed as rather unwanted features of the
theory, or these can indicate a strong instability of the dynamical system
that describes the cosmological evolution. Work is in progress towards the
latter possibility.


As a final task we shall try to compare the results of our analysis with
the recent Planck data \cite{planck}, by suitably choosing the
phenomenological EoS function. For the purposes of our analysis, we shall
express the observational indices as functions of the $e$-folding number
$N$. We shall consider three important observational indices, namely the
spectral index of primordial curvature perturbations $n_s$, the
scalar-to-tensor ration $r$ and the running of the spectral index $a_s$,
which as was evinced in Ref.~\cite{nojodineos1}, these are written in terms
of the $e$-folding number $N$ as follows,
\begin{align}
n_\mathrm{s} - 1 =& - 9 \rho(N) \tilde f(\rho (N)) \left( \frac{\tilde
f'(\rho (N))-2}{2\rho (N)
- \tilde f(\rho (N))}\right)^2
+\frac{6\rho (N)}{2\rho (N) - \tilde f(\rho (N))}
\left\{ \frac{ \tilde f(\rho (N))}{\rho (N)} \right. \nn
& \left. + \frac{1}{2} \left(\tilde f'(\rho (N))\right)^2
+ \tilde f'(\rho (N)) -\frac{5}{2} \frac{\tilde f(\rho (N)) \tilde f'(\rho
(N))}{\rho (N)}
+ \left(\frac{f(\rho)}{\rho(N)}\right)^2+\frac{1}{3} \frac{\rho
'(N)}{\tilde f(\rho (N))}
\right. \nonumber \\
& \left.
\times \left[\left( \tilde f'(\rho (N)) \right)^2 + \tilde f(\rho (N))
\tilde f''(\rho (N))
-2 \frac{ \tilde f(\rho (N)) \tilde f'(\rho (N))}{\rho (N)}
+ \left( \frac{ \tilde f(\rho (N))}{\rho (N)} \right)^2 \right] \right\} \,,
\label{eq:2.32} \\
r =& 24\rho (N) \tilde f(\rho (N))
\left( \frac{ \tilde f'(\rho (N))-2}
{2\rho (N) - \tilde f(\rho (N))}\right)^2 \,,
\label{eq:2.33} \\
\alpha_\mathrm{s} =& \rho (N) \tilde f(\rho (N)) \left( \frac{\tilde
f'(\rho (N))-2}
{2\rho (N) - \tilde f(\rho (N))}\right)^2 \left[ \frac{72\rho (N)}{2\rho (N)
- \tilde f(\rho (N))} J_1 \right. \nn
& \left. -54 \rho (N) \tilde f(\rho (N)) \left( \frac{\tilde f'(\rho (N))-2}
{2\rho (N) - \tilde f(\rho (N))}\right)^2 -\frac{1}{\tilde f'(\rho
(N))-2}J_2 \right] \,,
\label{eq:2.34}
\end{align}
where the detailed functional form of $J_1$ and $J_2$ is given in the
appendix. In order to have a qualitative idea of the behavior of the
observational indices in terms of the phenomenological EoS of
Eq.~(\ref{pesoparadigmfinal}), we shall study a not very sophisticated
model, but simple nevertheless, that leads to a Type IV singularity.
Suppose that the EoS is given by Eq.~(\ref{pesoparadigmfinal}), with
$f(\rho )=A \rho ^{\alpha}$ and in order to proceed we have to express the
function $f(\rho )$, in terms of the $e$-folding number $N$. This can be
easily done, since the scale factor in this case is equal to,
\begin{equation}\label{scalespecific}
a(t)=a_0 \e^{\frac{\rho ^{1-\alpha}}{3(1-\alpha)A}}\, ,
\end{equation}
so the effective energy density as a function of the $e$-folding number $N$
is given by,
\begin{equation}\label{reff}
\rho
=\left(3(1-\alpha)A\right)^{\frac{1}{1-\alpha}}N^{\frac{1}{1-\alpha}}\, .
\end{equation}
Recall that the function $f(\rho )$ can generate a Type IV singular
evolution when $0<\alpha<\frac{1}{2}$, as we demonstrated below
Eq.~(\ref{presstest}) and therefore, the fraction $f(\rho )/\rho $ can be
chosen to satisfy the constraint,
\begin{equation}\label{sda}
\frac{f(\rho )}{\rho }\ll 1\, .
\end{equation}
In this case, the observational indices can be very much simplified, and
can be approximated by the following expressions \cite{nojodineos1},
\begin{equation}\label{insapprox}
n_s\simeq 1-6\frac{f(\rho )}{\rho (N)}\, , \quad r\simeq 24\frac{f(\rho
)}{\rho (N)}\, , \quad \alpha_s=-9\left(\frac{f(\rho )}{\rho
(N)}\right)^2\, .
\end{equation}
Combining Eqs.~(\ref{reff}) and (\ref{insapprox}), with the constraint
(\ref{sda}) holding true, the resulting approximate expressions of the
observational indices read,
\begin{equation}\label{defrre}
n_s\simeq 1-\frac{2}{N (1-\alpha )}\, , \quad r\simeq \frac{8}{N (1-\alpha
)},{\,}{\,}\alpha_s\simeq -\frac{1}{N^2 (1-\alpha )^2}\, .
\end{equation}
The 2015 Planck report \cite{planck}, restricts the values of the
observational indices as follows,
\begin{equation}\label{planckconstr}
n_s=0.9644\pm 0.0049\, , \quad r<0.10\, , \quad a_s=-0.0057\pm 0.0071\, ,
\end{equation}
which can be relaxed if someone assumes a scale dependence of the scalar
and tensor spectral indices. By using the values $(N,\alpha)=(60,1/20)$,
the observational indices become equal to,
\begin{equation}\label{onsberf}
n_s\simeq 0.96491\, , \quad r\simeq 0.1403\, , \quad a_s=-0.000307\, .
\end{equation}
Therefore, there is concordance with the spectral index of primordial
curvature perturbations, but the scalar-to-tensor ratio and the associated
running of the spectral index constraints are not satisfied. However, in
principle concordance can be achieved, if a more sophisticated model is
considered, instead of the simple model we used here, just for expositional
purposes.

\section{Conclusions}

In this letter we investigated various phenomenological EoS models that can
generate a Type IV singular cosmological evolution. The motivation for
considering phenomenological EoS models is coming from the fact that the
Universe's EoS seems to be non-constant and it may slightly cross the
phantom
divide at late-time. In addition, the Type IV singularity is the mildest
possibility of singular evolutions, since it is of non-crushing type, and
therefore the Hawking-Penrose theorems \cite{hawkingpenrose} are satisfied
for this sort of singularity. However, as we evinced, the slow-roll
parameters might become singular at the singularity points, for specific
values of the free parameters that govern the cosmological evolution. The
singularity of the slow-roll parameters might indicate some sort of
instability of the dynamical system that the cosmological evolution
equations constitute. This singularity of the slow-roll parameters was also
observed in \cite{noo1,noo2,noo3} when evolution is governed by scalars,
so a repeating pattern seems to underlie
this feature, which by no chance is accidental. This issue needs to be
further scrutinized, and in general all the consequences of finite time
singularities should be fully understood in the context of classical
cosmology. This is owing to the fact that singularities, and especially the
crushing singularities, appearing in classical theories are usually
indicators of an underlying theoretical framework yet to be found.
Therefore, the ``mild'' singularities might be some classical link between
the quantum and the classical theory of gravity (apart from Type IV
singularity where quantum effects should not play an essential role).

An interesting possibility that we did not address in this letter is to
study a Type IV singular evolution caused by two distinct phenomenological
EoS dark fluids, or even couple one fluid with dark matter. This study
however exceeds the introductory purposes of this paper and is deferred to
a future work.

Finally, a comment is in order: The generalized models we used are not by any means just mathematical constructions, since such EoS models are typical for acceleration of the Universe, since their EoS parameter is around the value $-1$. Their specific form can be imposed in principle, by a specific evolution of the Universe, therefore these EoS are physically  motivated by the early/current status of the evolution of the Universe.

\section*{Acknowledgments}

This work is supported in part by the JSPS Grant-in-Aid for Scientific
Research (C) \# 23540296 (S.N.) and in part by MINECO (Spain), projects
FIS2010-15640 and FIS2013-44881 (S.D.O.)

\section*{Appendix}

Here we provide the detailed functional form of the parameters $J_1$ and
$J_2$ that appear in the expressions giving the observational indices,
namely in Eq.~(\ref{eq:2.34}), as functions of the effective energy
density, and the $e-$folding number. The detailed form of these is,
\begin{align}
J_1 \equiv& \frac{\tilde f(\rho (N))}{\rho (N)} + \frac{1}{2}
\left(\tilde f'(\rho (N))\right)^2 + \tilde f'(\rho (N)) 
-\frac{5}{2} \frac{\tilde f(\rho (N)) \tilde f'(\rho (N))}{\rho (N)}
+ \left(\frac{\tilde f(\rho (N))}{\rho (N)}\right)^2 \nonumber+\frac{1}{3}
\frac{\rho '(N)}{\tilde f(\rho (N))} \nn
&
\times \left[\left(\tilde f'(\rho (N))\right)^2 + \tilde f(\rho (N)) \tilde
f''(\rho (N))
-2 \frac{\tilde f(\rho (N)) \tilde f'(\rho (N))}{\rho (N)} + \left(
\frac{\tilde f(\rho (N))}{\rho (N)}
\right)^2 \right]\,, \label{eq:2.35} \\
J_2 \equiv &
\frac{45}{2} \frac{\tilde f(\rho (N))}{\rho (N)} \left(\tilde f'(\rho (N))
- \frac{1}{2} \frac{\tilde f(\rho (N))}{\rho (N)}\right) 
+ 18\left(\frac{\tilde f(\rho (N))}{\rho (N)}\right)^{-1} \left\{
\left(\tilde f'(\rho (N))-\frac{1}{2} \frac{ \tilde f(\rho (N))}{\rho
(N)}\right)^2 \right. \nn
& \left. + \left( \tilde f'(\rho (N))-\frac{1}{2} \frac{\tilde f(\rho
(N))}{\rho (N)}\right)^3 \right\} 
-9\left( \tilde f'(\rho (N)) - \frac{1}{2} \frac{\tilde f(\rho (N))}{\rho
(N)}\right)^2
-45 \tilde f'(\rho (N)) + 9\frac{\tilde f(\rho (N))}{\rho (N)}
\nonumber \\
&
+3 \left(4 \tilde f'(\rho (N)) -7\frac{ \tilde f(\rho (N))}{\rho (N)}
+2\right)
\left\{
-\frac{3}{2}\left(\tilde f'(\rho (N)) -\frac{1}{2}\frac{\tilde f(\rho
(N))}{\rho (N)}\right)
+ \left(\frac{\tilde f(\rho (N))}{\rho (N)}\right)^{-2}
\frac{\rho '(N)}{\rho (N)} \right. \nn
& \times \left[ \left( \tilde f'(\rho (N))\right)^2 + \tilde f(\rho (N))
\tilde f''(\rho (N)) 
\left.
- 2 \frac{ \tilde f(\rho (N)) \tilde f'(\rho (N))}{\rho (N)} + \left(
\frac{\tilde f(\rho (N))}{\rho (N)}
\right)^2 \right] \right\}
\nonumber \\
& +\left(\frac{ \tilde f(\rho (N))}{\rho (N)}\right)^{-2}
\left\{
-\frac{3}{2} \left(\frac{\tilde f(\rho (N))}{\rho (N)}\right)
\left(\frac{\rho '(N)}{\rho (N)}\right)
\left[
3\left( \tilde f'(\rho (N)) \right)^2
+2 \tilde f(\rho (N)) \tilde f''(\rho (N)) \right. \right. \nn
& \left. -\frac{11}{2}\frac{ \tilde f(\rho (N)) \tilde f'(\rho (N))}{\rho
(N)}
+\frac{5}{2} \left( \frac{\tilde f(\rho (N))}{\rho (N)} \right)^2
\right]
\nonumber \\
& \left.
+ \left(\frac{\rho ''(N)}{\rho (N)}\right)
\left[
\left( \tilde f'(\rho (N)) \right)^2 + \tilde f(\rho (N)) \tilde f''(\rho
(N))
-2\frac{ \tilde f(\rho ) \tilde f'(\rho )}{\rho (N)}
+ \left( \frac{ \tilde f(\rho )}{\rho } \right)^2
\right]
\right.
\nonumber \\
& \left.
+ \left(\frac{\rho '(N)}{\rho (N)}\right)^2
\left[ \left(3\tilde f'(\rho (N)) \tilde f''(\rho (N)) + \tilde f(\rho (N))
\tilde f'''(\rho (N)) \right) \rho (N)
-3\left( \tilde f'(\rho (N)) \right)^2 \right. \right. \nn
& \left. \left. - 3 \tilde f(\rho (N)) \tilde f''(\rho (N))
+6\frac{\tilde f(\rho (N)) \tilde f'(\rho (N))}{\rho (N)}
-3\left( \frac{\tilde f(\rho (N))}{\rho (N)} \right)^2
\right]
\right.
\,.
\label{eq:2.36}
\end{align}

\end{document}